\documentclass[11pt,a4paper]{article}
\usepackage{jheppub}
\pdfoutput=1

\usepackage{amsmath, amsfonts, amssymb}
\usepackage{color, graphicx, perpage, setspace}
\usepackage{hyperref}

\usepackage[english]{babel}
\usepackage[utf8]{inputenc}
\usepackage[T1]{fontenc}

\MakePerPage[2]{footnote}

\begin{document}
\hfill UMN--TH--3905/19\\

\title{A Holographic Perspective on the Axion Quality Problem}

\author[a,b]{Peter Cox,} 
\author[c]{Tony Gherghetta,}
\author[c]{Minh D. Nguyen}

\affiliation[a]{School of Physics, The University of Melbourne, Victoria 3010, Australia}
\affiliation[b]{Kavli Institute for the Physics and Mathematics of the Universe (WPI), UTIAS, The University of Tokyo, Kashiwa, Chiba 277-8583, Japan}
\affiliation[c]{School of Physics and Astronomy, University of Minnesota, Minneapolis, Minnesota 55455, USA}

\emailAdd{peter.cox@unimelb.edu.au}
\emailAdd{tgher@umn.edu}
\emailAdd{nguy1642@umn.edu}

\abstract{The axion provides a compelling solution to the strong CP problem as well as a candidate for the dark matter of the universe. However, the axion solution relies on the spontaneous breaking of a global $U(1)_{PQ}$ symmetry, which is also explicitly violated by quantum gravity. To preserve the axion solution, gravitational violations of the $U(1)_{PQ}$ symmetry must be suppressed to sufficiently high order. We present a simple, geometric solution of the axion quality problem by modelling the axion with a bulk complex scalar field in a slice of AdS$_5$, where the $U(1)_{PQ}$ symmetry is spontaneously broken in the bulk but explicitly broken on the UV brane. By localising the axion field towards the IR brane, gravitational violations of the PQ symmetry on the UV brane are sufficiently sequestered. This geometric solution is holographically dual to 4D strong dynamics where the global $U(1)_{PQ}$ is an accidental symmetry to sufficiently high order.}

\maketitle


\section{Introduction}

The strong CP problem remains one of the most intriguing puzzles in the Standard Model (SM). The strong interaction  CP-violating parameter ${\bar \theta}= \theta_{QCD} - {\rm arg\,det\,}M_q$, where $\theta_{QCD}$ is the QCD vacuum angle and $M_q$ the bare quark mass matrix, is constrained to be $\bar\theta \lesssim 10^{-10}$ by the non-observation of the neutron electric dipole moment. This experimental fact requires a correlation between two seemingly different parts of the theory, the QCD vacuum angle and the phases in the quark mass matrix. Furthermore, given that there is no anthropic argument for why $\bar\theta$ is so small, it strongly suggests that there should be a dynamical solution to the strong CP problem.

The favoured solution to the strong CP problem is the Peccei-Quinn (PQ) mechanism~\cite{Peccei:1977hh}. A spontaneously broken global $U(1)_{PQ}$ symmetry gives rise to a Nambu-Goldstone boson that is identified as the axion~\cite{Weinberg:1977ma,Wilczek:1977pj}. Non-perturbative QCD effects explicitly break the symmetry and generate a nonzero axion potential. The minimum of this potential occurs at a value for which the axion field exactly cancels the source of CP violation, $\bar\theta$. This solution is both simple and elegant, and as an added bonus the axion could even be responsible for the dark matter of the universe~\cite{Preskill:1982cy,Abbott:1982af,Dine:1982ah}. This makes the axion one of the most motivated new particles of beyond the Standard Model physics.

However, besides the explicitly violating non-perturbative QCD effects, the axion solution requires that the $U(1)_{PQ}$ global symmetry be preserved to an extraordinarily high degree, otherwise the minimum of the axion potential no longer cancels $\bar\theta$.  In particular, violations of the $U(1)_{PQ}$ global symmetry arise from quantum gravity where it is well known that global symmetries are not preserved. Thus, Planck-suppressed operators which violate the $U(1)_{PQ}$ symmetry must be forbidden to sufficiently high order~\cite{Holman:1992us,Kamionkowski:1992mf,Barr:1992qq,Ghigna:1992iv}.\footnote{It is possible that quantum gravity only violates global symmetries non-perturbatively via gravitational instantons, in which case higher-dimensional operators are  exponentially suppressed~\cite{Alonso:2017avz}.} Some of the possible ways to address this so-called axion quality problem include considering a composite axion~\cite{Kim:1984pt,Choi:1985cb} and realising the Peccei-Quinn symmetry as an accidental symmetry~\cite{Redi:2016esr,Lillard:2017cwx, Lillard:2018fdt,Gavela:2018paw}, gauging the Peccei-Quinn symmetry~\cite{Fukuda:2017ylt,Fukuda:2018oco,Ibe:2018hir} or lowering the scale of spontaneous breaking which occurs in heavy axion models~\cite{Rubakov:1997vp,Berezhiani:2000gh,Hook:2014cda,Fukuda:2015ana,Gherghetta:2016fhp,Dimopoulos:2016lvn}.

In this paper, we present a holographic solution of the axion quality problem in a slice of AdS$_5$~\cite{Randall:1999ee}. Specifically, using the AdS/CFT correspondence~\cite{Maldacena:1997re}, we model the $U(1)_{PQ}$ global symmetry of some underlying strong dynamics as a 5D $U(1)_{PQ}$ gauge symmetry that is spontaneously broken in the bulk. A massive, complex scalar field charged under the gauge symmetry corresponds to a PQ-charged operator ${\cal O}$ of dimension $\Delta$. Both the explicit and spontaneous breaking of the global $U(1)_{PQ}$ symmetry are modelled by a vacuum expectation value for the bulk scalar field, with the former sourced by UV boundary terms that explicitly violate the $U(1)_{PQ}$ global symmetry. 
The axion is then identified with the pseudoscalar fluctuations around this background, with a profile that depends on the arbitrary dimension $\Delta$. Varying $\Delta$ allows the axion to be localised towards the IR brane where it is naturally sequestered from explicit gravitational violations of the Peccei-Quinn symmetry on the UV brane. A lower limit of at least $\Delta\gtrsim 10$ can then be derived to sufficiently suppress UV sources of explicit breaking for an axion decay constant $F_a\gtrsim 10^9$~GeV. This 5D model gives a simple, geometric interpretation of the axion quality problem and is dual to 4D solutions that invoke composite axions with an accidental $U(1)_{PQ}$ symmetry.

Our axion solution builds on previous work that considered axions propagating in an extra dimension. Axions in a flat extra dimension were first studied in \cite{Dienes:1999gw}. An axion that arises from a higher-dimensional gauge field in a warped dimension was discussed in \cite{hep-ph/0308024}. Our solution is related but has the important difference that the symmetry is spontaneously broken by a bulk scalar vacuum expectation value (corresponding to an operator of finite dimension), as opposed to IR brane boundary conditions (corresponding to an operator of infinite dimension). Furthermore, the holographic dual of our 5D setup closely resembles 4D composite axion models.
Other realisations of axions in a two-throat warped geometry were given in \cite{Flacke:2006ad}. The equations of motion we derive for the pseudoscalar sector generalise similar results first obtained in the context of AdS/QCD for the QCD pseudoscalar sector~\cite{DaRold:2005vr}. For pions in QCD, the global chiral symmetry is spontaneously broken by a dimension three operator corresponding to a bilinear fermion condensate with explicit breaking by the quark masses on the UV boundary. Our result generalises the QCD pseudoscalar solution to arbitrary operator dimension $\Delta$; furthermore, we explicitly show that the Nambu-Goldstone mode remains massless, even in the presence of a UV brane, when there is no explicit violation of the global symmetry.

The outline of this paper is as follows. In Section 2 we present the 5D model in a slice of AdS$_5$ and derive the equations of motion and boundary conditions. In Section 3 we present the massless axion solutions which correspond to the spontaneous breaking of a PQ-charged operator with arbitrary dimension $\Delta$.
The massive axion solutions are given in Section 4, where an explicit source of PQ violation is added on the UV brane. Two realistic composite axion models are presented in Section 5 corresponding to placing QCD either on the UV brane or in the bulk. As expected, for QCD on the UV brane it is difficult to sequester the gravitational violation, while for QCD in the bulk the axion quality problem can be addressed. Our conclusion is given in Section 6.


\section{5D setup} \label{sec:5Dsetup}

To model a global $U(1)_{PQ}$ symmetry of some underlying strong dynamics, we will consider a scalar field, $\Phi$, charged under a $U(1)_{PQ}$ gauge symmetry in a slice of $AdS_5$. In conformal coordinates, the 5D metric takes the form
\begin{equation}
  ds^2=A^2(z)\left(dx^2+dz^2\right)\equiv g_{MN} dx^M dx^N\,,
\end{equation}
where the 5D coordinates are denoted $x^M=(x^\mu,z)$,  and $A(z) = 1/(kz)$ is the warp factor with $k$ the $AdS_5$ curvature scale~\cite{Randall:1999ee}. A UV (IR) brane is located at $z_{UV}\, (z_{IR})$. The relevant 5D action is given by
\begin{align} \label{eq:5D_action}
  S&=2\int^{z_{IR}}_{z_{UV}}d^5x\, \sqrt{-g} \left( -\frac{1}{4g_5^2}F^{MN}F_{MN} - \frac{1}{2} \big(D^M\Phi^\dagger\big)\big(D_M\Phi\big) - \frac{1}{2}m_\Phi^2\Phi^\dagger\Phi \right. \notag \\
  &\hspace{4cm}\left.-\frac{1}{2g_5^2\xi}\left( g^{\mu\nu}\partial_\mu V_\nu + \xi A^{-3}\partial_z\left(AV_z\right) - \xi g_5^2\eta^2a \right)^2 \right) \notag \\
  &- \int d^4x\, \sqrt{-g_4} \, U(\Phi) \,,
\end{align}
where the complex scalar field is parametrised as $\Phi=\eta\, e^{ia}$, and the $U(1)$ gauge field is denoted by $V_M=(V_\mu,V_z)$ with $g_5$ the 5D gauge coupling and $D_M=\partial_M-iV_M$. The $U(\Phi)$ are boundary potentials on the UV and IR branes whose form will be specified later. 
We work in $R_\xi$ gauge (with gauge parameter $\xi$), where the vector and scalar modes decouple. 

\subsection{Background solution}

We restrict to the case where the backreaction of the scalar $\Phi$ on the metric can be neglected\footnote{This requires $|(\partial_z\eta)^2-m_\Phi^2\eta^2|\ll 12k^2M_5^3$, where $M_5$ is the 5D Planck mass which satisfies $M_P^2 \simeq M_5^3/k$, with the reduced Planck mass $M_P=2.4\times 10^{18}$~GeV.}; the equation of motion for the $z$-dependent scalar vacuum expectation value, $\eta(z)$, is then
\begin{equation} \label{eq:eta-eom}
  \partial_z\left(A^3\partial_z\eta\right) - m_\Phi^2A^5\eta = 0 \,,
\end{equation}
with the boundary condition,
\begin{equation} \label{eq:eta-bc}
  \partial_z\eta \mp \frac{A}{2} \frac{dU}{d\eta} \Bigg|_{z_{UV},\,z_{IR}} =0\,,
\end{equation}
where the upper\,(lower) signs correspond to $z_{UV}$\,($z_{IR}$). 
The equation of motion (\ref{eq:eta-eom}) has the general solution
\begin{equation} \label{eq:bulk-scalar}
  \eta(z) = k^{3/2} \left( \lambda\, (kz)^{4-\Delta} + \sigma\, (kz)^{\Delta} \right) \,,
\end{equation}
where $\Delta>2$ is related to the bulk scalar mass according to $m_\Phi^2=\Delta(\Delta-4)\,k^2$~\cite{Witten:1998qj}. 
The dimensionless coefficients $\lambda$ and $\sigma$ are fixed by the boundary conditions in eq.~\eqref{eq:eta-bc}, and the boundary potentials are assumed to have the following form
\begin{align}
  U_{UV}(\Phi) &= (-\ell_{UV}k^{5/2}\Phi + \text{h.c.} ) + b_{UV} k\, \Phi^\dagger\Phi \,, \label{eq:UV-potential} \\
  U_{IR}(\Phi) &= \frac{\lambda_{IR}}{k^2}\left(\Phi^\dagger\Phi-k^3v_{IR}^2\right)^2 \,, \label{eq:IR-potential}
\end{align}
where $\ell_{UV},b_{UV},\lambda_{IR}$ and $v_{IR}$ are real dimensionless coefficients. 
Note that the linear term in \eqref{eq:UV-potential} explicitly breaks the $U(1)_{PQ}$ symmetry on the UV brane. 
Solving the boundary conditions and taking the limit $z_{IR} \gg z_{UV}$ with $\Delta>4$ one obtains
\begin{align}
  \lambda &= \frac{\ell_{UV}}{\Delta-4+b_{UV}}(kz_{UV})^{\Delta-4} \,, \label{eq:lambda} \\
  \sigma &= \sqrt{v_{IR}^2-\frac{\Delta}{2\lambda_{IR}}}\,(kz_{IR})^{-\Delta}\equiv \sigma_0\,(kz_{IR})^{-\Delta} \,.
\end{align}
Notice that $\sigma$ is suppressed by $(kz_{IR})^{-\Delta}$, while $\lambda$ is $\mathcal{O}(1)$.
When $\ell_{UV}$=0 sub-leading terms need to be kept and one finds in this case that $\lambda$ is also suppressed:
\begin{equation} \label{eq:lambda-sigma}
  \lambda = \frac{\Delta-b_{UV}}{\Delta-4+b_{UV}}(kz_{UV})^{2\Delta-4}\sigma \,.
\end{equation}
If $b_{UV}=0$ then $\lambda$ in \eqref{eq:lambda-sigma} is fixed in terms $\Delta$ and $\sigma$; a nonzero value of $b_{UV}$ allows $\lambda$ to be independently varied.

Using the AdS/CFT correspondence, we can interpret the above 5D setup in terms of a dual strongly interacting 4D conformal field theory (CFT). 
The presence of the UV and IR branes correspond to explicit and spontaneous breaking of the conformal symmetry respectively, with the latter giving rise to a mass-gapped theory. 
The scalar $\Phi$ is identified with a PQ-charged operator $\mathcal{O}$, with dimension $\Delta$, in the dual strongly coupled sector.
Furthermore, $\sigma$ is identified with a condensate $\langle\mathcal{O}\rangle$, and $\ell_{UV}$ with turning on a source for $\mathcal{O}$.
This source explicitly breaks $U(1)_{PQ}$, and for $\Delta>4$ corresponds to breaking the global symmetry by a Planck-suppressed operator. 


\subsection{Pseudoscalar sector}

We are interested in the pseudoscalar spectrum, and in particular the lightest mode which will be identified with the axion. 
Varying the action (\ref{eq:5D_action}) with respect to $V_z$ and $a$ one obtains the following equations of motion
\begin{align} 
  \label{eq:bulkVz}
  A\,\Box V_z + g_5^2A^3\eta^2\left(\partial_z a - V_z\right) + \xi A\partial_z\left(A^{-1}\partial_z\left(AV_z\right) - g_5^2A^2\eta^2a\right) &= 0 \,, \\
  \label{eq:bulka}
  A^3\eta^2\,\Box a + \partial_z\left(A^3\eta^2\left(\partial_z a - V_z\right)\right) + \xi A^2\eta^2\left(\partial_z\left(AV_z\right) - g_5^2A^3\eta^2a\right) &= 0 \,,
\end{align}
where $\Box\equiv\eta^{\mu\nu}\partial_\mu\partial_\nu$ and $\eta^{\mu\nu}=\text{diag}(-,+,+,+)$. 
We solve the equations of motion by performing the KK expansion,
\begin{align} \label{eq:KK-expansion}
  a(x^\mu,z) &= \sum_{n=0}^\infty f_{a}^{(n)}(z) a^{(n)}(x^\mu) \,, \\
  V_z(x^\mu,z) &= \sum_{n=0}^\infty f_{V_z}^{(n)}(z) a^{(n)}(x^\mu) \,,
\end{align}
where $a^{(n)}(x^\mu)$ satisfies $\Box a^{(n)} =m_n^2 a^{(n)}$.
Note that $a$ and $V_z$ are expanded in terms of the same set of 4D modes; if $a$ and $V_z$ were expanded in different 4D modes then the orthogonality of the eigenvectors of the d'Alembertian transforms \eqref{eq:bulkVz} and \eqref{eq:bulka} into four separate equations, whose solution does not satisfy the boundary condition \eqref{eq:Vz-bdry} (to be imposed below).

The boundary conditions are
\begin{align} 
  \pm \frac{2}{g_5^2} \left( A\eta^{\mu\nu}\partial_\mu V_\nu + \xi\left( \partial_z\left(AV_z\right) - g_5^2A^3\eta^2a \right)\right) \delta V_z \bigg|_{z_{UV},z_{IR}} &= 0 \,, \label{eq:Vz-bdry-general} \\
  \left( \pm2A^3\eta^2\left(\partial_za - V_z\right) -A^4\frac{\delta U}{\delta a} \right)\delta a \bigg|_{z_{UV},z_{IR}} &= 0\,, \label{eq:a-bdry-general} \\
  \pm\frac{2}{g_5^2}A\left(\partial_zV_\mu - \partial_\mu V_z\right)\delta V^\mu \bigg|_{z_{UV},z_{IR}} &= 0 \,, \label{eq:Vmu-bdry-general} 
\end{align}
where we have included the boundary condition obtained from varying the action with respect to $V_\mu$, since this does not decouple from $V_z$. 
It is important to note that the 5D gauge symmetry imposes further restrictions on the boundary conditions that can be used to satisfy the above conditions~\cite{hep-th/0604121}. 
In order to have a well-defined 5D gauge transformation, one cannot impose Dirichlet conditions for $V_z$ and either of $V_\mu$ or $a$ on the same boundary since doing so would constrain the form of gauge transformations in the bulk. 

To satisfy eqs.~\eqref{eq:Vz-bdry-general}-\eqref{eq:Vmu-bdry-general} we impose the following boundary conditions on the fields
\begin{align}
  V_\mu \Big|_{z_{UV}} &= 0 \,, \qquad &\partial_zV_\mu \Big|_{z_{IR}} =0 \,, \label{eq:Vmu-bdry} \\
  \xi\left( \partial_z\left(AV_z\right) - g_5^2A^3\eta^2a \right) \Big|_{z_{UV}} &= 0 \,, \qquad &V_z \Big|_{z_{IR}} =0 \,, \label{eq:Vz-bdry} \\
  \pm2A^3\eta^2\left(\partial_za - V_z\right) -A^4\frac{\delta U}{\delta a} \Bigg|_{z_{UV},z_{IR}} &= 0 \,. \label{eq:a-bdry}
\end{align}
These boundary conditions then restrict the 5D gauge symmetry on the boundaries, where the gauge transformation parameter $\alpha(x^\mu,z)$ must satisfy
\begin{align} \label{eq:restrictedGT}
  \partial_\mu \alpha\Big|_{z_{UV}} &= 0 \,, \qquad \partial_z \alpha\Big|_{z_{IR}} = 0 \,, \notag \\
  \partial_z(A\partial_z\alpha)-g_5^2A^3\eta^2\alpha \Big|_{z_{UV}} &= 0 \,,
\end{align}
for a general infinitesimal 5D gauge transformation
\begin{equation}
  V_M \to V_M + \partial_M \alpha(x^\mu,z) \,, \qquad
  a \to a + \alpha(x^\mu,z) \,.
\end{equation}
The Dirichlet condition on $V_\mu$ at $z=z_{UV}$ therefore restricts the gauge symmetry to a global symmetry on the UV brane, and ensures that there is no massless vector mode in the spectrum (i.e. the global $U(1)_{PQ}$ symmetry in the dual 4D CFT is not gauged). 
This also determines the UV boundary condition for $V_z$.
On the other hand, we want the gauge symmetry to be preserved on the IR boundary and so impose a Neumann condition for $V_\mu$ at $z=z_{IR}$; this also fully determines the IR boundary conditions for $V_z$ and $a$.
The reason for this choice is that we are interested in the spontaneous breaking of the $U(1)_{PQ}$ symmetry by the scalar $\Phi$, which is dual to spontaneous breaking by an operator of dimension $\Delta$ in the 4D CFT. 
If we were to instead impose a Dirichlet condition for $V_\mu$ it would correspond to spontaneous breaking by the infinite dimension operator associated with the IR brane. 
In this case one recovers the model of \cite{hep-ph/0308024} in the limit that the scalar field $\Phi$ is decoupled. Furthermore, since our choice of boundary conditions preserves the gauge symmetry in the IR, all explicit sources of $U(1)_{PQ}$ violation are confined to the UV brane. 


\section{Massless axion} \label{sec:massless-axion}

In this section we first look for solutions that describe an exactly massless axion ($m_0=0$). 
We therefore require that there is no source of explicit $U(1)_{PQ}$ breaking in the UV by taking $\ell_{UV}=0$. 
We also assume $\lambda=0$, which follows from imposing the condition $b_{UV}=\Delta$ in \eqref{eq:lambda-sigma}; we will comment on the case $\lambda\neq0$ in section~\ref{sec:nonzero-lambda}.

\begin{figure}[t]
  \centering
  \includegraphics[height=4.65cm]{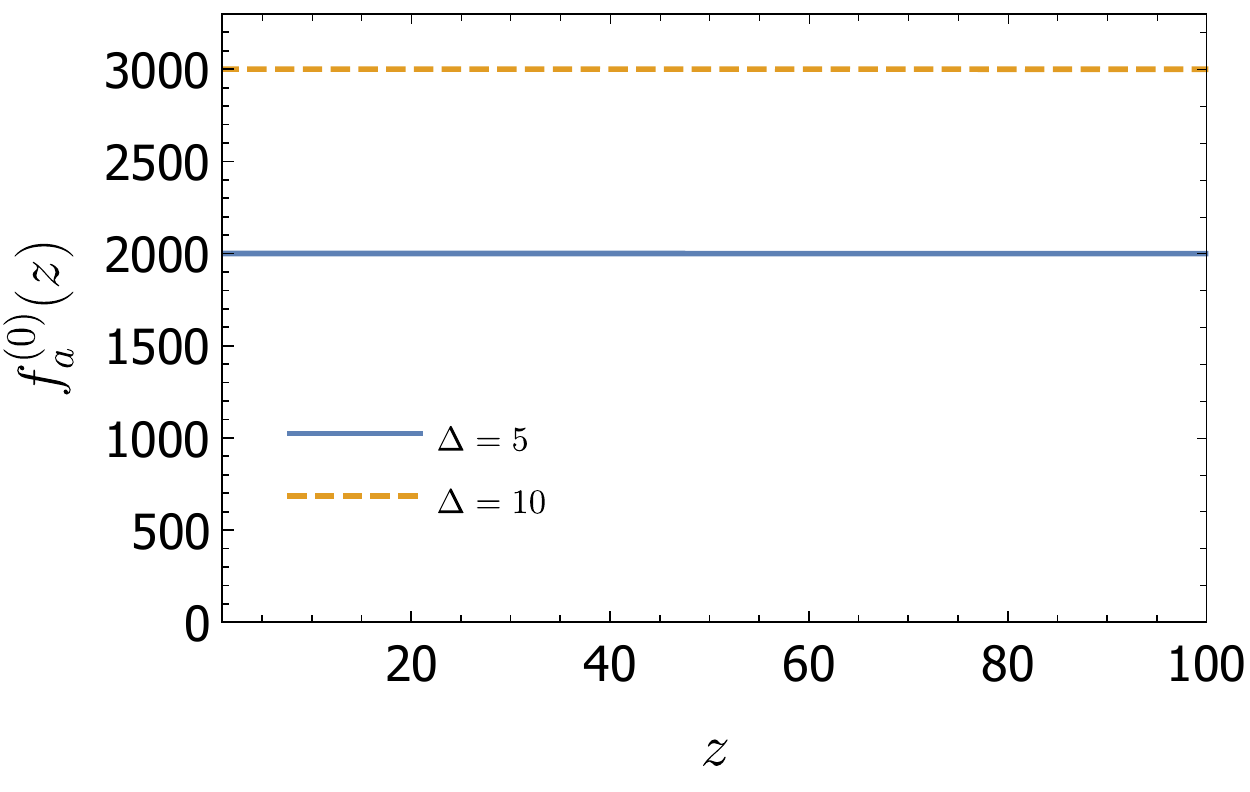}
  \includegraphics[height=4.65cm]{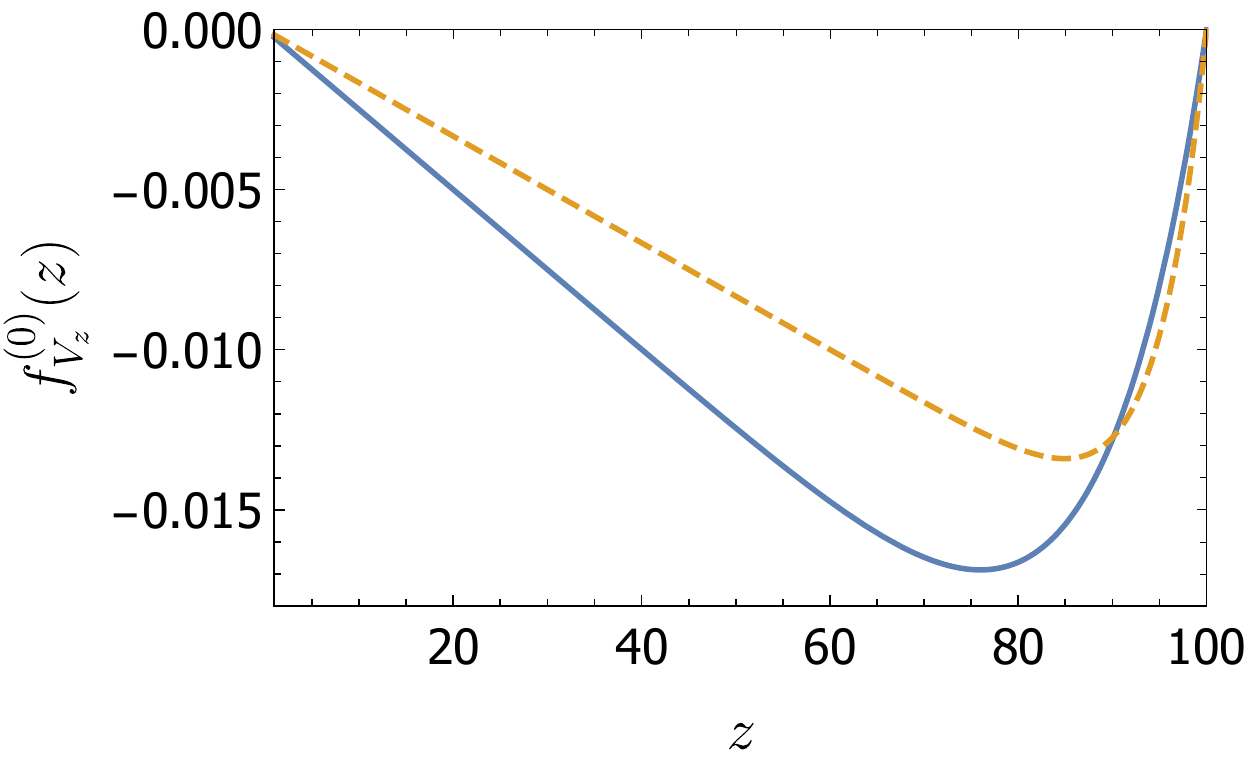}
  \caption{Normalised massless zero mode profiles \eqref{eq:massless-profile} with $k=1$, $kz_{UV}=1$, $kz_{IR}=100$, $g_5\sqrt{k}=1$, and $\sigma_0=0.1$.}
  \label{fig:massless-profiles}
\end{figure}

First, it is useful to define the new fields
\begin{align}
  \label{eq:chi-def}
  \chi &= \partial_za-V_z \,, \\
  \label{eq:zeta-def}
  \zeta &= \frac{1}{A} \left( \partial_z\left(AV_z\right) - g_5^2A^3\eta^2a \right)\,.
\end{align}
Notice that $\chi$ is gauge invariant. 
In terms of these new fields the equations of motion, \eqref{eq:bulkVz} and \eqref{eq:bulka}, reduce to a coupled first order system for the massless modes in the KK expansion
\begin{align} \label{eq:eom-firstorder}
  g_5^2A^3\eta^2f_\chi^{(0)} + \xi A\partial_zf_\zeta^{(0)} &= 0 \,, \notag \\
  \partial_z\left(A^3\eta^2f_\chi^{(0)}\right) + \xi A^3\eta^2f_\zeta^{(0)} &= 0 \,.
\end{align}
This has the general solution (for $\lambda=0$):
\begin{align} \label{eq:chi-zeta-solutions}
  f_\chi^{(0)}(z) &= -  (kz)^{2-\Delta}\left( c_1\, I_{\frac{1}{\Delta}-1}  \left( {\scriptstyle g_5\sqrt{k}\frac{\sigma}{\Delta} (kz)^\Delta } \right) + c_2\, I_{1-\frac{1}{\Delta}} \left( {\scriptstyle g_5\sqrt{k}\frac{\sigma}{\Delta} (kz)^\Delta } \right) \right) \,, \notag \\
  f_\zeta^{(0)}(z) &= \frac{g_5}{\xi}k^{3/2}\sigma \, (k z) \left( c_1\, I_{\frac{1}\Delta} \left( {\scriptstyle g_5\sqrt{k}\frac{\sigma}{\Delta} (kz)^\Delta } \right) + c_2\, I_{-\frac{1}\Delta} \left( {\scriptstyle g_5\sqrt{k}\frac{\sigma}{\Delta} (kz)^\Delta } \right) \right) \,,
\end{align}
where $I_\alpha(x)$ is the modified Bessel function of the first kind, and $c_{1,2}$ are dimensionless constants. 
The boundary conditions in eqs.~\eqref{eq:Vz-bdry} and \eqref{eq:a-bdry} require that at least one of the $f_\chi^{(0)}$ or $f_\zeta^{(0)}$ vanish on each boundary. 
This is enough to enforce $c_1=c_2=0$ such that $f_\chi^{(0)}$ and $f_\zeta^{(0)}$ vanish everywhere.
Despite this, the solution is non-trivial when expressed back in terms of $V_z$ and $a$; solving eqs.~\eqref{eq:chi-def} and \eqref{eq:zeta-def} we obtain the profiles
\begin{align} \label{eq:massless-profile}
  f_{V_z}^{(0)}(z) &= g_5\sqrt{k}\sigma (kz)^\Delta \left( c_3\, I_{\frac{1}{\Delta}-1} \left( {\scriptstyle g_5\sqrt{k}\frac{\sigma}{\Delta} (kz)^\Delta } \right) + c_4\, I_{1-\frac{1}{\Delta}} \left( {\scriptstyle g_5\sqrt{k}\frac{\sigma}{\Delta} (kz)^\Delta } \right) \right) \,, \notag \\
  f_{a}^{(0)}(z) &= z \left( c_3\, I_{\frac{1}\Delta} \left( {\scriptstyle g_5\sqrt{k}\frac{\sigma}{\Delta} (kz)^\Delta } \right) + c_4\, I_{-\frac{1}\Delta} \left( {\scriptstyle g_5\sqrt{k}\frac{\sigma}{\Delta} (kz)^\Delta } \right) \right) \,,
\end{align}
where $c_3,c_4$ are dimensionless constants.
Imposing the remaining IR boundary condition, $V_z(z_{IR})=0$, gives
\begin{equation}
  c_3 = - \frac{ I_{1-\frac{1}{\Delta}} \left( {\scriptstyle g_5\sqrt{k}\frac{\sigma}{\Delta} (kz_{IR})^\Delta } \right) } { I_{\frac{1}{\Delta}-1} \left( {\scriptstyle g_5\sqrt{k}\frac{\sigma}{\Delta} (kz_{IR})^\Delta } \right) } c_4 \,.
\end{equation}
The last integration constant is fixed by canonically normalising the profiles, with the relevant part of the 5D action given by
\begin{equation} \label{eq:kinetic-action}
  S \supset 2 \int_{z_{UV}}^{z_{IR}} d^5x \, \left( \frac{1}{2g_5^2}AV_z \Box V_z + \frac{1}{2}A^3\eta^2 a \Box a \right) \,.
\end{equation}
The resulting profiles for $f_a^{(0)}$ and $f_{V_z}^{(0)}$ are shown in figure~\ref{fig:massless-profiles}. 

Further insight can be gained by looking at the approximate profiles obtained by expanding the Bessel functions in eq.~\eqref{eq:massless-profile} for small argument. 
For large $z_{IR}$, this will be a very good approximation away from the IR brane since $\sigma=\sigma_0(kz_{IR})^{-\Delta}$.
It remains a good approximation in the IR provided that $\sigma_0<1$; we will always assume this is the case since it also ensures that the backreaction of the scalar can be neglected. 
The approximate profiles are
\begin{align} \label{eq:approx-massless-profile}
  f_{V_z}^{(0)}(z) &\simeq \frac{-1}{2\sigma_0\sqrt{\Delta-1}} \frac{z}{z_{IR}}\left( g_5^2 k\sigma_0^2 \left(1-\left(\frac{z}{z_{IR}}\right)^{2(\Delta-1)}\right) + \mathcal{O}(\sigma_0^4) \right) \,, \notag \\
  f_a^{(0)}(z) &\simeq \frac{z_{IR}}{\sigma_0} \sqrt{\Delta-1} \left(1 + \frac{g_5^2 k\sigma_0^2}{4\Delta(\Delta-1)}\left( \frac{(\Delta-1)^2}{2\Delta-1} + \frac{z^2}{z_{IR}^2}\left( \left(\frac{z}{z_{IR}}\right)^{2(\Delta-1)} - \Delta\right)\right)  + \mathcal{O}(\sigma_0^4) \right) \,,
\end{align}
where in the normalisation constant we have also taken $z_{IR}\gg z_{UV}$.
Notice that the leading term in the $f_{a}^{(0)}$ profile is constant, while the solution for $f_{V_z}^{(0)}$ is approximately given by $f_{V_z}^{(0)}\propto A^{-1}$ in the UV.

\subsection{Global \texorpdfstring{$U(1)_{PQ}$}{U(1)PQ} symmetry}

The global $U(1)_{PQ}$ symmetry that acts as a shift symmetry for the axion is realised as a subset of the 5D gauge transformations. 
Specifically, consider the subset of gauge transformations where the gauge parameter depends only on $z$ and takes the same form as $f_a^{(0)}(z)$:
\begin{equation}
  \alpha(x^\mu,z) = \alpha_0 \, f_a^{(0)}(z) \,,
\end{equation}
where $\alpha_0$ is an arbitrary constant. 
One can easily check that this form also satisfies the boundary conditions in eq.~\eqref{eq:restrictedGT}. 
Since $f_{V_z}^{(0)}(z)=\partial_z f_a^{(0)}(z)$, these gauge transformations act as a shift symmetry on the 4D axion mode; specifically,
\begin{equation} \label{eq:shift-symmetry}
  \quad a^{(0)}(x^\mu) \to a^{(0)}(x^\mu) + \alpha_0 \,.
\end{equation}
This is just the action of the global $U(1)_{PQ}$ symmetry in the 4D effective theory.

\subsection{Comment on \texorpdfstring{$\lambda\neq0$}{nonzero lambda}} \label{sec:nonzero-lambda}

So far, we have restricted ourselves to the assumption that $\lambda=0$. 
A na\"ive application of the standard AdS/CFT dictionary might suggest that this is a necessary condition to obtain a massless mode, since a non-zero $\lambda$ is usually identified with turning on a source for the corresponding operator in the dual CFT. 
However, when $\ell_{UV}=0$ the UV boundary condition in eq.~\eqref{eq:eta-bc} relates $\lambda$ and $\sigma$ in \eqref{eq:lambda-sigma}; $\lambda$ is then associated with a source for the operator $\mathcal{O}^\dagger\mathcal{O}$~\cite{Witten:2001ua} and so does not explicitly break $U(1)_{PQ}$. 
The scaling with $z_{UV}$ in \eqref{eq:lambda-sigma} also shows that $\lambda$ should be identified with a source for an operator of dimension $2\Delta$.
The situation changes once $\ell_{UV}\neq0$, as can be seen from eq.~\eqref{eq:lambda}.
This now describes a CFT with a non-zero source for the operator $\mathcal{O}$, which explicitly breaks $U(1)_{PQ}$.
We will consider this case in section~\ref{sec:massive-axion}.

From the point of view of the 5D theory, there is no explicit breaking of the $U(1)_{PQ}$ global symmetry when $\ell_{UV}=0$, even for non-zero $\lambda$.
In fact, one can quite easily see that there is still a massless mode in the spectrum, since $\chi=\zeta=0$ is a solution to eq.~\eqref{eq:eom-firstorder} for any value of $\lambda$. 
The only difference is that it is no longer straightforward to solve eqs.~\eqref{eq:chi-def} and \eqref{eq:zeta-def} to obtain expressions for the profiles (except in the limit $g_5\to0$). 


\section{Massive axion} \label{sec:massive-axion}

Global symmetries are expected to be violated by quantum gravity. 
This in general presents a significant hurdle to axion solutions to the strong CP problem, since the stringent upper bound on $\bar\theta$ requires $U(1)_{PQ}$ to be an extremely good approximate global symmetry.
This problem can be addressed if the global symmetry has its origin as a gauge symmetry in higher dimensions~\cite{hep-ph/0308024}, since this severely restricts possible sources of explicit $U(1)_{PQ}$ symmetry breaking. 
Within the current setup, the 5D gauge symmetry restricts global $U(1)_{PQ}$ breaking to two possible sources: (i) terms localised on the UV brane, where the gauge symmetry is reduced to a global symmetry ($\partial_\mu\alpha(x^\mu,z_{UV})=0$); (ii) bulk terms that transform as a total derivative under gauge transformations, such as a Chern-Simons term. 
It is important to note that although the global $U(1)_{PQ}$ symmetry that acts as a shift symmetry on the axion in eq.~\eqref{eq:shift-symmetry} is explicitly broken on the UV brane, there remains a 5D gauge symmetry with a gauge parameter that satisfies $\alpha(x^\mu,z_{UV})=0$. 
Furthermore, in the case of a bulk Chern-Simons term, fermions must be added on the IR brane to cancel the localised gauge anomaly.

For now, let us focus on UV-localised sources and look for a solution that describes a massive axion in the presence of $U(1)_{PQ}$ breaking effects from Planck-suppressed operators. 
To achieve this we include a UV boundary potential for $\Phi$ that explicitly breaks the global $U(1)_{PQ}$ symmetry. 
The leading effects will come from a term linear in $\Phi$; this can also be easily understood from the point of view of the dual CFT, where it corresponds to adding a source term for the operator $\mathcal{O}$ that spontaneously breaks $U(1)_{PQ}$. 

It is straightforward to see that the inclusion of such a boundary term will give rise to a mass for the axion. 
Rewriting the linear term in eq.~\eqref{eq:UV-potential} in terms of $a$ gives
\begin{equation}
  U_{UV}(\Phi) \supset -2\ell_{UV}k^{5/2}\eta\cos(a) = -2\ell_{UV}k^{5/2}\eta\left(1- \frac{1}{2}a^2 + \ldots \right) \,.
\end{equation}
More precisely, the above potential modifies the boundary condition in eq.~\eqref{eq:a-bdry} such that there is no longer a massless mode in the spectrum.

\begin{figure}[t]
  \centering
  \includegraphics[height=4.8cm]{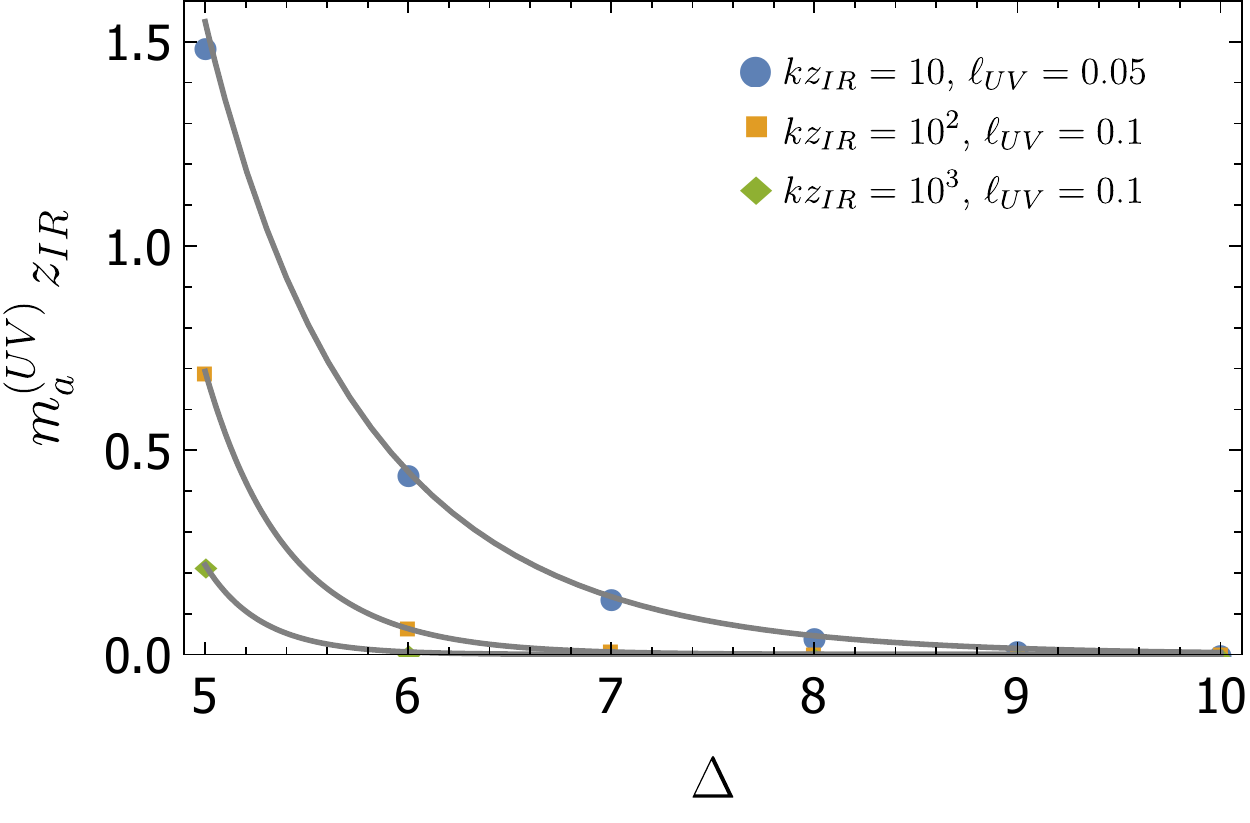}
  \includegraphics[height=4.8cm]{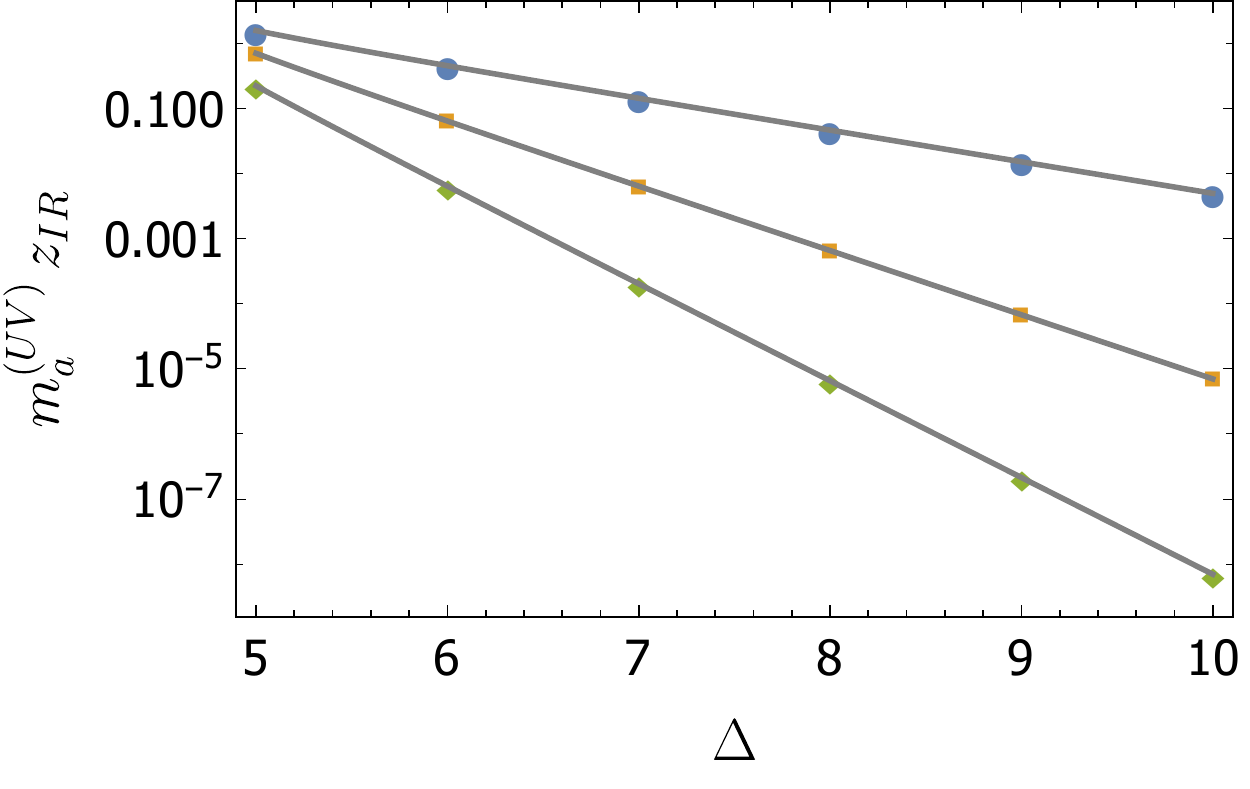}
  \caption{Axion mass relative to the compositeness scale ($z_{IR}^{-1}$), with $kz_{UV}=1$, $\sigma_0=0.1$ and $b_{UV}=0$. The curves correspond to the analytic solution for $g_5=0$ \eqref{eq:axion-mass}, while the points are obtained by numerically solving the equations of motion with $g_5\sqrt{k}=1$.}
  \label{fig:axion-mass}
\end{figure}

For massive modes and non-zero $\lambda$ it is no longer straightforward to solve the equations of motion in general; however, an analytic solution can be obtained perturbatively in $g_5\sqrt{k}$. 
Note that since the $g_5$-dependent terms in the equations of motion are also proportional to $\eta^2(z)$, this expansion is expected to provide a good approximation even for relatively large values of $g_5$, as $\eta(z)\lesssim1$ if the scalar backreaction can be neglected. 
For our purposes it is sufficient to work at leading order in $g_5$; the equations of motion then simplify significantly, since from eq.~\eqref{eq:kinetic-action} one can see that $V_z$ must vanish at zeroth order. 
Eq.~\eqref{eq:bulka} can then be solved to obtain the $f_a^{(n)}$ profile,
\begin{equation}
  f_a^{(n)}(z) = \frac{\sqrt{k}(k z)^2}{\eta(z)} \left(\frac{m_n}{k}\right)^{2-\Delta} \left( d_1 J_{\Delta-2}(m_n z) + d_2 Y_{\Delta-2}(m_n z) \right) \,,
\end{equation}
where $d_{1,2}$ are dimensionless constants.  
Given that we are predominantly interested in the lightest mode, for which we expect $m_0z_{IR}\ll1$ (assuming $\Delta>4$), it is useful to expand the Bessel functions for small argument to obtain the approximate axion profile,
\begin{multline} \label{eq:approx-massive-profile-unnormalised}
  f_a^{(0)}(z) \simeq \frac{\sqrt{k}(k z)^{4-\Delta}}{\eta(z)} \left( \frac{d_1}{2^\Delta\Gamma(\Delta)} \left( 4(\Delta-1) - (m_0 z)^2 \right) (kz)^{2(\Delta-2)} \right. \\
  \left. - \frac{d_2 \Gamma(\Delta-3)}{2^{4-\Delta}\pi} \left(\frac{m_0}{k}\right)^{2(2-\Delta)} \left( 4(\Delta-3) + (m_0 z)^2 \right) \right) \,.
\end{multline}
Imposing the IR boundary condition, $\partial_z {f_a^{(0)}}|_{z_{IR}}=0$, yields
\begin{multline}
  d_2 = -d_1\frac{\pi\, 4^{2-\Delta}}{\Gamma(\Delta)\Gamma(\Delta-3)} (m_0z_{IR})^{2(\Delta-2)} \\
  \times \frac{\lambda (\Delta-1) ((m_0z_{IR})^2-4(\Delta-2)) + \sigma_0 (m_0z_{IR})^2 (kz_{IR})^{\Delta-4}}{\lambda(m_0z_{IR})^2-\sigma_0 (\Delta-3)(kz_{IR})^{\Delta-4}((m_0z_{IR})^2+4(\Delta-2))} \,.
\end{multline}
The axion mass $(m_a^{(UV)}\equiv m_0)$ is then determined by the UV boundary condition:
\begin{align} \label{eq:axion-mass}
  (m_a^{(UV)} z_{IR})^2 &= \frac{4\ell_{UV}}{\sigma_0}  \frac{\lambda (\Delta-1)(\Delta-2) (kz_{IR})^{4-\Delta}}{\ell_{UV} +2(\Delta-2)\sigma_0(z_{UV}/z_{IR})^\Delta}~, \\
  &\simeq \frac{4\ell_{UV}}{\sigma_0} \frac{(\Delta-1)(\Delta-2)}{\Delta-4+b_{UV}} \left(\frac{z_{IR}}{z_{UV}}\right)^{4-\Delta} \,.
  \label{eq:axion-mass-approx}
\end{align}
In the first line we have taken the $z_{IR} \gg z_{UV}$ limit for simplicity, but have also kept the leading $\ell_{UV}$-independent term. 
The mass is proportional to $\ell_{UV}$ and vanishes in the absence of explicit breaking in the UV (so far we have not included the coupling to QCD). 
In the second line above we have used eq.~\eqref{eq:lambda} and taken the limit $z_{IR} \gg z_{UV}$. 
The factor $(z_{IR}/z_{UV})^{4-\Delta}$ can be understood from the dual theory as being due to the RG running from the UV scale down to the confinement scale ($\sim z_{IR}^{-1}$), and is consistent with explicit breaking by an operator of dimension $\Delta$. 
This results in a significant suppression of the axion mass when $z_{IR} \gg z_{UV}$ and $\Delta$ is large. 
This is shown in figure~\ref{fig:axion-mass}, where we also compare with the mass obtained by numerically solving the equations of motion with $g_5\sqrt{k}=1$. 
Notice that eq.~\eqref{eq:axion-mass} continues to be a good approximation when $g_5\sim1$, since $\eta(z)\lesssim0.1$. 

\begin{figure}[t]
  \centering
  \includegraphics[width=0.7\textwidth]{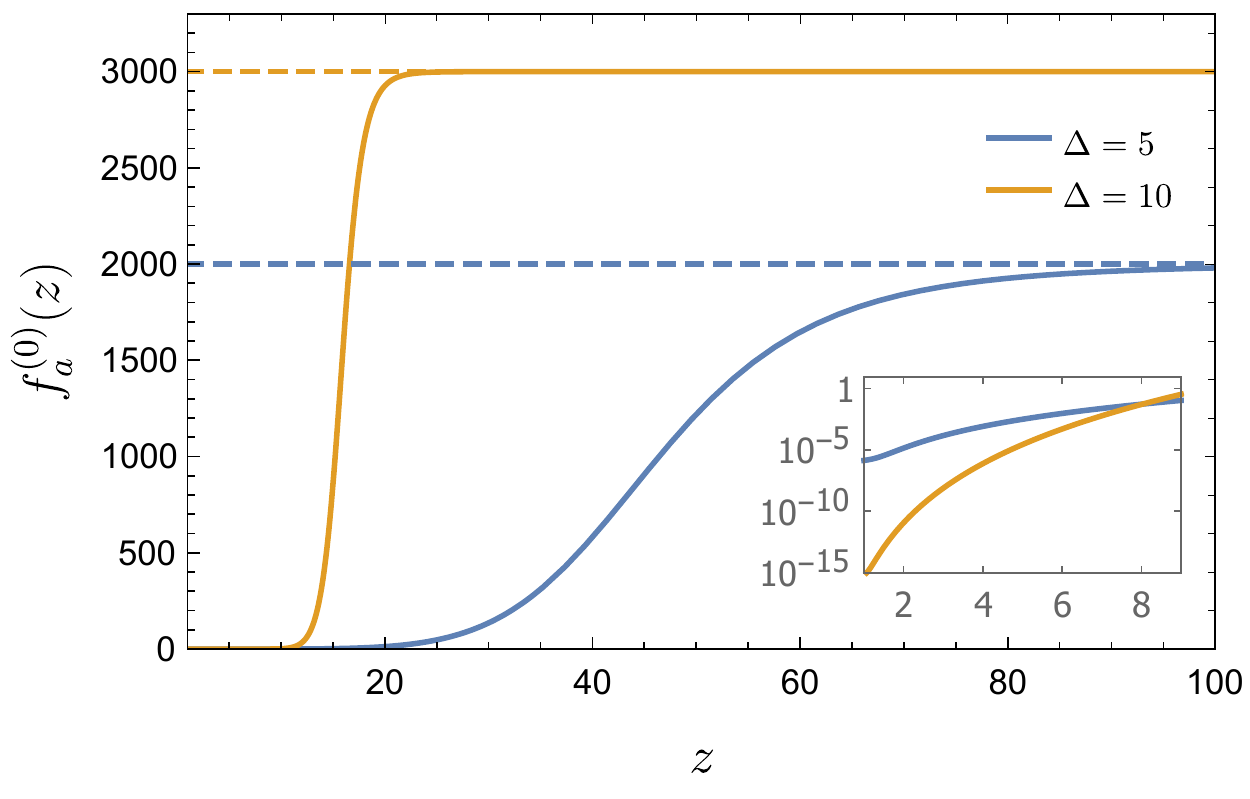}
  \caption{Normalised axion profile with explicit $U(1)_{PQ}$ breaking UV boundary term. The solid lines show the approximate profile in eq.~\eqref{eq:approx-massive-profile}, while the dashed lines show the massless ($\ell_{UV}=\lambda=0$) solution \eqref{eq:massless-profile} for comparison. We fixed $k=1$, $kz_{UV}=1$, $kz_{IR}=100$, $\sigma_0=0.1$, $\ell_{UV}=0.1$, and $b_{UV}=0$.}
  \label{fig:massive-profile}
\end{figure}

\begin{figure}[t]
  \centering
  \includegraphics[height=4.65cm]{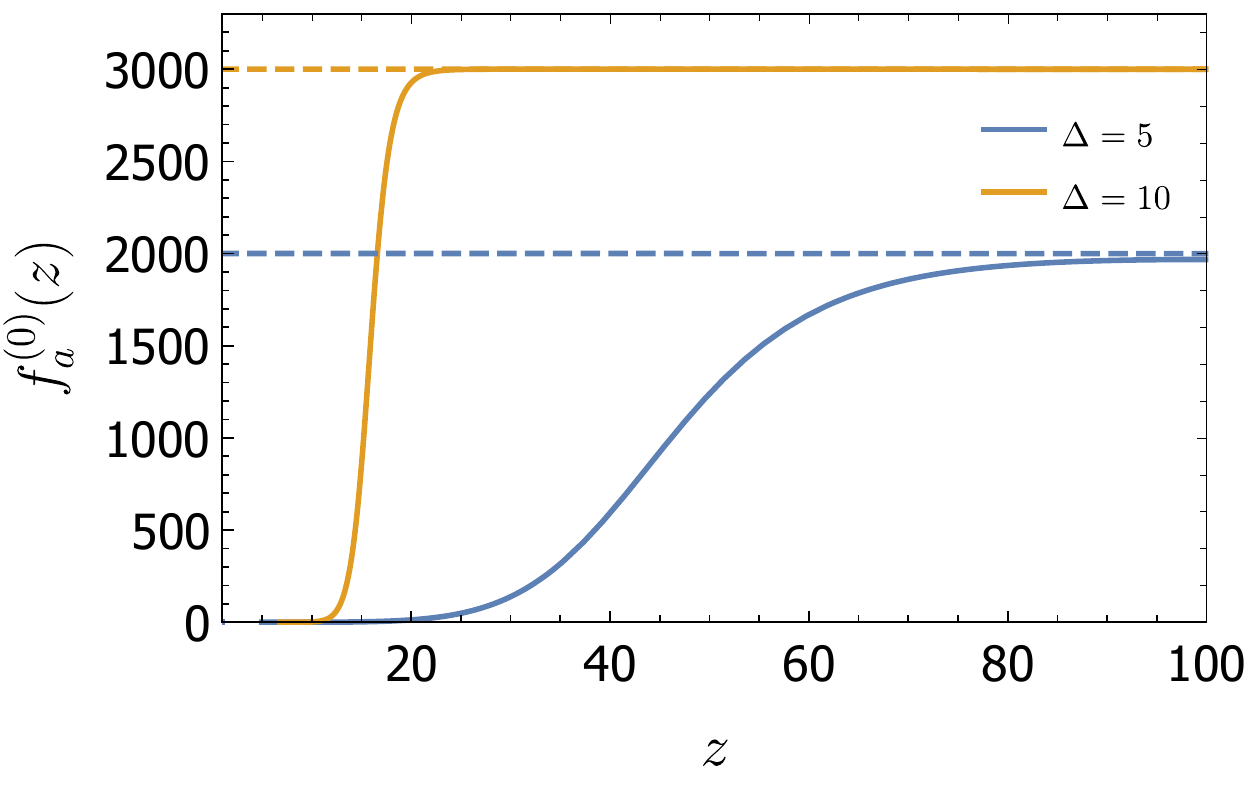}
  \includegraphics[height=4.65cm]{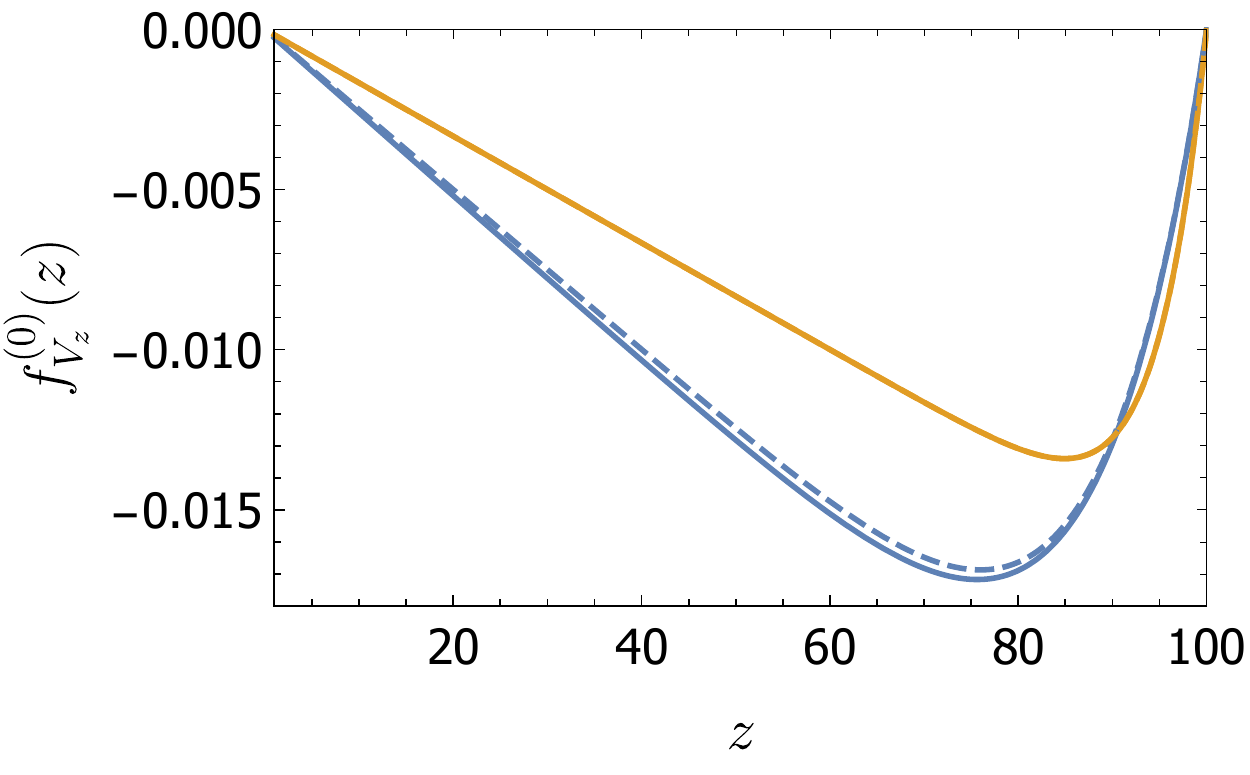}
  \caption{Same as figure~\ref{fig:massive-profile}, except with $g_5\sqrt{k}=1$ and the solid lines have been obtained by numerically solving the equations of motion.}
  \label{fig:massive-profile-numerical}
\end{figure}

We fix the remaining constant $d_1$ by canonically normalising the kinetic term in eq.~\eqref{eq:kinetic-action}. 
This requires the solution for $V_z$ at $\mathcal{O}(g_5\sqrt{k})$, which is obtained by solving \eqref{eq:bulkVz}:
\begin{equation}
  f_{V_z}^{(n)}(z) = g_5\sqrt{k} z \left( d_3 J_0\left(\frac{m_n}{\sqrt{\xi}}z\right) + d_4 Y_0\left(\frac{m_n}{\sqrt{\xi}}z\right) \right) \,,
\end{equation}
where $d_{3,4}$ are dimensionless constants. 
However, to satisfy the boundary conditions \eqref{eq:Vz-bdry} and \eqref{eq:a-bdry} requires $d_3=d_4=0$. 
The leading term in $V_z$ is then $\mathcal{O}(g_5^2k)$, as was previously found for the massless solution in \eqref{eq:approx-massless-profile}, and can be neglected to the order we are working. Substituting  \eqref{eq:approx-massive-profile-unnormalised} into eq.~\eqref{eq:kinetic-action} then gives 
\begin{equation}
  d_1 = \frac{2^{\Delta-2} \Gamma(\Delta)}{\sqrt{\Delta-1}} (k z_{IR})^{1-\Delta} \,.
\end{equation}
Putting everything together, the approximate axion profile, valid when $z_{IR} \gg z_{UV}$, is 
\begin{equation} \label{eq:approx-massive-profile}
   f_a^{(0)}(z) \simeq z_{IR}\frac{k^{3/2}}{\eta(z)}\sqrt{\Delta-1}\left(\frac{z}{z_{IR}}\right)^\Delta \left[ 1 + \frac{2\lambda(\Delta-2)(k z_{UV})^\Delta (kz)^{2(2-\Delta)}}{\ell_{UV}+2\sigma_0(\Delta-2)(z_{UV}/z_{IR})^\Delta}\right] \,.
\end{equation}
The axion profile is plotted in figure~\ref{fig:massive-profile}. 
Notice that, in contrast to the massless case, the profile becomes highly suppressed in the UV, particularly for large $\Delta$.
This feature will play an important role when constructing composite axion models in the following section. 
Figure~\ref{fig:massive-profile-numerical} shows the profiles for both $f_a^{(0)}$ and $f_{V_z}^{(0)}$ with $g_5\sqrt{k}=1$, obtained by numerically solving the equations of motion. 
The $f_a^{(0)}$ profile closely matches the perturbative solution in figure~\ref{fig:massive-profile}, while $f_{V_z}^{(0)}$ remains largely unchanged from the massless case. 

\section{Composite axion models} \label{sec:models}

In this section we show how the 5D solution obtained in the previous section can be used to construct holographic descriptions of composite axion models~\cite{Kim:1984pt,Choi:1985cb}. 
This requires introducing the usual coupling between the axion and QCD ($SU(3)_c$).
There are in principle two ways to do this, and each corresponds to a different class of composite axion models:
\begin{itemize}
  \item $SU(3)_c$ localised on the UV brane 
  \item $SU(3)_c$ in the bulk
\end{itemize}
In the following sections we discuss each of the above models, and in particular show that only the latter can provide a solution to the axion quality problem.

\subsection{\texorpdfstring{$SU(3)_c$}{SU(3)c} on the UV brane}

Here the Standard Model fields are localised on the UV brane. 
One can then add either an additional Higgs doublet or additional coloured fermions to construct a DFSZ-\cite{Zhitnitsky:1980tq,Dine:1981rt} or KSVZ-type~\cite{Kim:1979if,Shifman:1979if} model on the UV brane.  
In either case, the effective action for the axion takes the form
\begin{equation}
  \mathcal{S}_{eff} = \int d^4x \left( \frac{1}{2} a^{(0)} \left( \Box - m_a^2 \right) a^{(0)} + \frac{N g_s^2}{32\pi^2} f_a^{(0)}(z_{UV}) a^{(0)} G\tilde{G} + \frac{Eg^2}{32\pi^2} f_a^{(0)}(z_{UV}) a^{(0)}  F\tilde{F} + \cdots \right) \,,
\end{equation}
where the anomaly coefficients $E$ and $N$ are determined by the charges of the UV localised fermions, and $g_s\,(g)$ is the QCD\,(QED) coupling. 
The axion decay constant is therefore determined by the value of the profile on the UV brane. 
Using eq.~\eqref{eq:approx-massive-profile} we obtain
\begin{multline}
  F_a = \frac{1}{f_a^{(0)}(z_{UV})} = \frac{z_{IR}^{-1}}{\sqrt{\Delta-1}} \left(\frac{z_{IR}}{z_{UV}}\right)^{\Delta} (\lambda(kz_{UV})^{4-\Delta} + \sigma_0(z_{UV}/z_{IR})^\Delta) \\
  \times \left[ 1 + \frac{2\lambda(\Delta-2)(k z_{UV})^{4-\Delta}}{\ell_{UV}+2\sigma_0(\Delta-2)(z_{UV}/z_{IR})^\Delta}\right]^{-1} \,.
\end{multline}
When $\ell_{UV}=0$ the decay constant is of order the IR scale, $z_{IR}^{-1}$ (recall that $\lambda \propto z_{IR}^{-\Delta}$ when $\ell_{UV}=0$, see \eqref{eq:lambda-sigma}). 
However, once explicit sources of $U(1)_{PQ}$ breaking are included the axion profile becomes highly suppressed in the UV, as shown in Figure~\ref{fig:massive-profile}. 
The effective decay constant for couplings to UV localised fields is then significantly larger than the UV scale, $z_{UV}^{-1}$, as it is enhanced by the factor $(z_{IR}/z_{UV})^{\Delta-1}$. 
This scaling can be understood as a consequence of partial compositeness in the dual theory, where the coupling arises via mixing between the composite operator of dimension $\Delta$ and an elementary scalar with mass of order $z_{UV}^{-1}$.

Since all UV boundary terms see the same effective axion decay constant, there is no way to suppress non-QCD sources of $U(1)_{PQ}$ breaking. 
Hence, this model cannot solve the axion quality problem.\footnote{This construction might still have an interesting application as a continuum limit of clockwork models~\cite{1704.07831}, since it allows for hierarchically different axion couplings on the UV and IR branes.}

\subsection{\texorpdfstring{$SU(3)_c$}{SU(3)c} in the bulk}

The second class of models involves enlarging the 5D gauge symmetry to $SU(3)_c \times U(1)_{PQ}$. 
We assume that the rest of the SM fields are confined to the UV brane. 
The axion coupling to $G\tilde{G}$ can be generated from the Chern-Simons term,
\begin{equation}\label{eq:Chern-Simons}
  -\frac{\kappa}{32\pi^2} \int_{z_{UV}}^{z_{IR}} d^5x\, \epsilon^{MNPQR} V_M G^a_{NP} G^a_{QR} \,,
\end{equation}
where $\kappa$ is a dimensionless constant and $\epsilon^{MNPQR}$ is the 5D Levi-Civita tensor density. 
Under a 5D gauge transformation, $V_M \to V_M + \partial_M\alpha$, this term is only invariant up to a total derivative, giving rise to boundary terms
\begin{equation} 
  \delta S = -\frac{\kappa}{32\pi^2} \left[ \int d^4x\, \alpha(x^\mu,z)\, \epsilon^{\mu\nu\rho\sigma}G^a_{\mu\nu}G^a_{\rho\sigma} \right]^{z_{IR}}_{z_{UV}} \,.
\end{equation}
The gauge parameter satisfies $\alpha(x^\mu,z_{UV})=0$ and so the $z=z_{UV}$ term above vanishes. 
However, in the IR there is a localised gauge anomaly that needs to be cancelled by adding appropriately charged fermions on the IR brane (one might therefore expect that $\kappa$ is quantised). 
In the effective action for the axion, this is equivalent to adding the term
\begin{equation} \label{eq:IR-ct}
  \frac{\kappa}{32\pi^2}\int d^4x\,a \, G \tilde{G} \bigg|_{z_{IR}} \,.
\end{equation}
Thus, the combined action \eqref{eq:Chern-Simons} and \eqref{eq:IR-ct} is invariant under the gauge transformation.
Integrating over $z$, and using the fact that the massless gluon profile is constant, we obtain the effective action for the axion,
\begin{equation}
  \mathcal{S}_{eff} = \int d^4x \left( \frac{1}{2} a^{(0)} \left( \Box - m_a^2 \right) a^{(0)} + \frac{g_s^2}{32\pi^2F_a} a^{(0)} G\tilde{G} \right) \,,
\end{equation}
where from eqs.~\eqref{eq:Chern-Simons} and \eqref{eq:IR-ct} the axion decay constant is
\begin{equation}
  \frac{1}{F_a} = \kappa \left( f_{a}^{(0)}(z_{IR}) - \int_{z_{UV}}^{z_{IR}} dz \, f_{V_z}^{(0)}(z) \right) \,.
\end{equation}
Substituting the profile in eq.~\eqref{eq:approx-massive-profile} gives
\begin{equation} \label{eq:decay_constant}
  F_a \simeq \frac{1}{\kappa} \frac{\sigma_0}{\sqrt{\Delta-1}} z_{IR}^{-1}  \,,
\end{equation}
and the axion decay constant is of order the IR scale. 
We have confirmed numerically that this also remains the case when $g_5\sqrt{k}$ is $\mathcal{O}(1)$.

On the other hand, axion couplings to any additional, UV localised sources of $U(1)_{PQ}$ breaking are highly suppressed in this model. 
This is again a consequence of the fact that the $f_a^{(0)}(z)$ profile is IR localised and becomes highly suppressed in the UV when $\Delta$ is large.\footnote{Note that if there were additional bulk fields charged under $U(1)_{PQ}$ these could spoil the suppression if their 5D masses corresponded to operators of lower dimension. On the other hand, any contributions to the axion potential from additional UV-localised PQ-charged fields will be suppressed.} 
In addition, $V_z$ can only appear in the UV boundary action in the gauge invariant combination $F_{\mu z}=\partial_\mu V_z -\partial_z V_\mu$. 
This is due to the fact that $V_z$, unlike $a$, still transforms non-trivially on the UV brane under 5D gauge transformations, $V_z\rightarrow V_z +\partial_z \alpha|_{z_{UV}}$, as $\partial_z\alpha|_{z_{UV}}\neq0$.
Therefore, since $V_z$ must be derivatively coupled, UV localised sources of explicit breaking only generate a potential for the axion through their coupling to $a$, which is suppressed.

The bulk $SU(3)_c\times U(1)_{PQ}$ model therefore provides a realistic, holographic description of a composite axion that can solve the axion quality problem. For large $\Delta$ the effects of explicit breaking sources in the UV can be sufficiently suppressed, while the axion decay constant that determines the QCD contribution to the axion potential is only weakly dependent on $\Delta$. 
For a given decay constant, there is then a minimum critical value $\Delta_c$ needed to address the axion quality problem. 

\begin{figure}
    \centering
    \includegraphics[height=5.2cm]{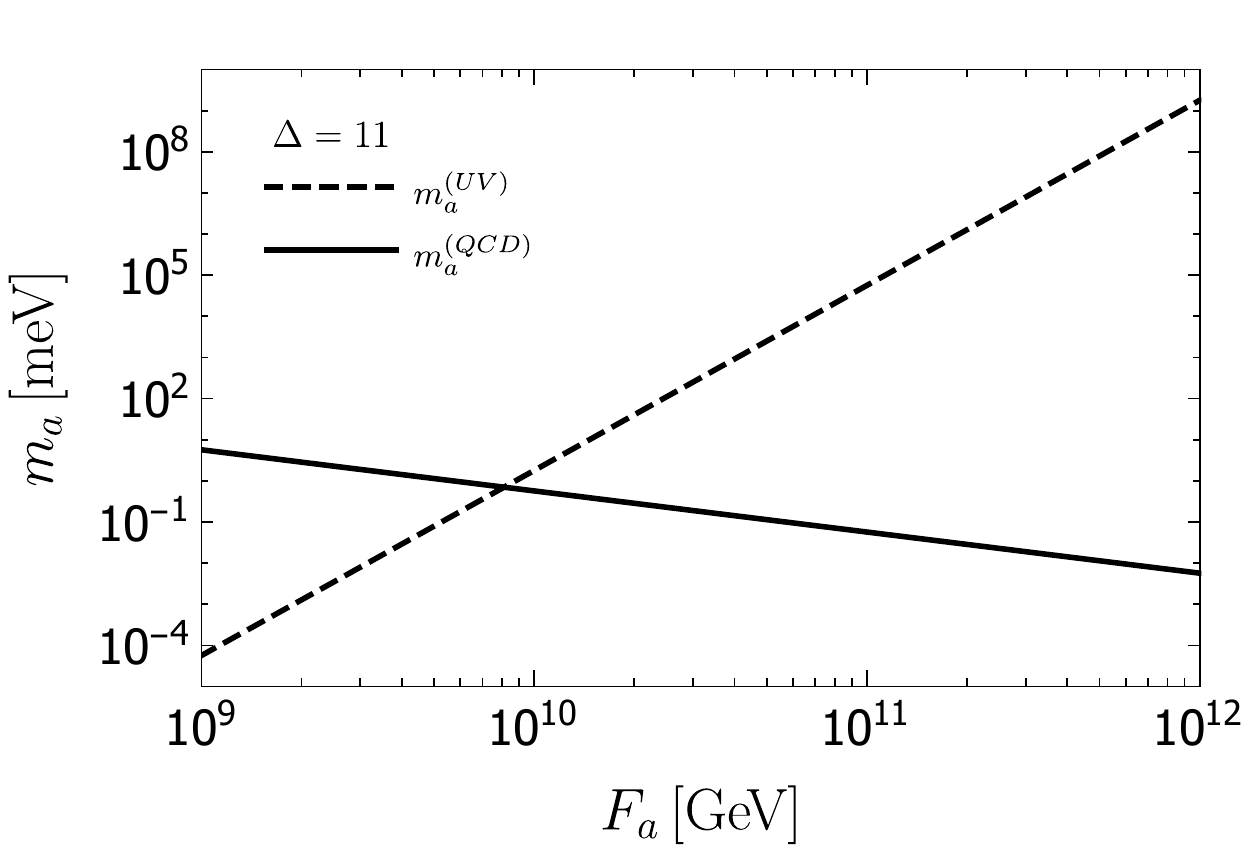}
    \includegraphics[height=5.2cm]{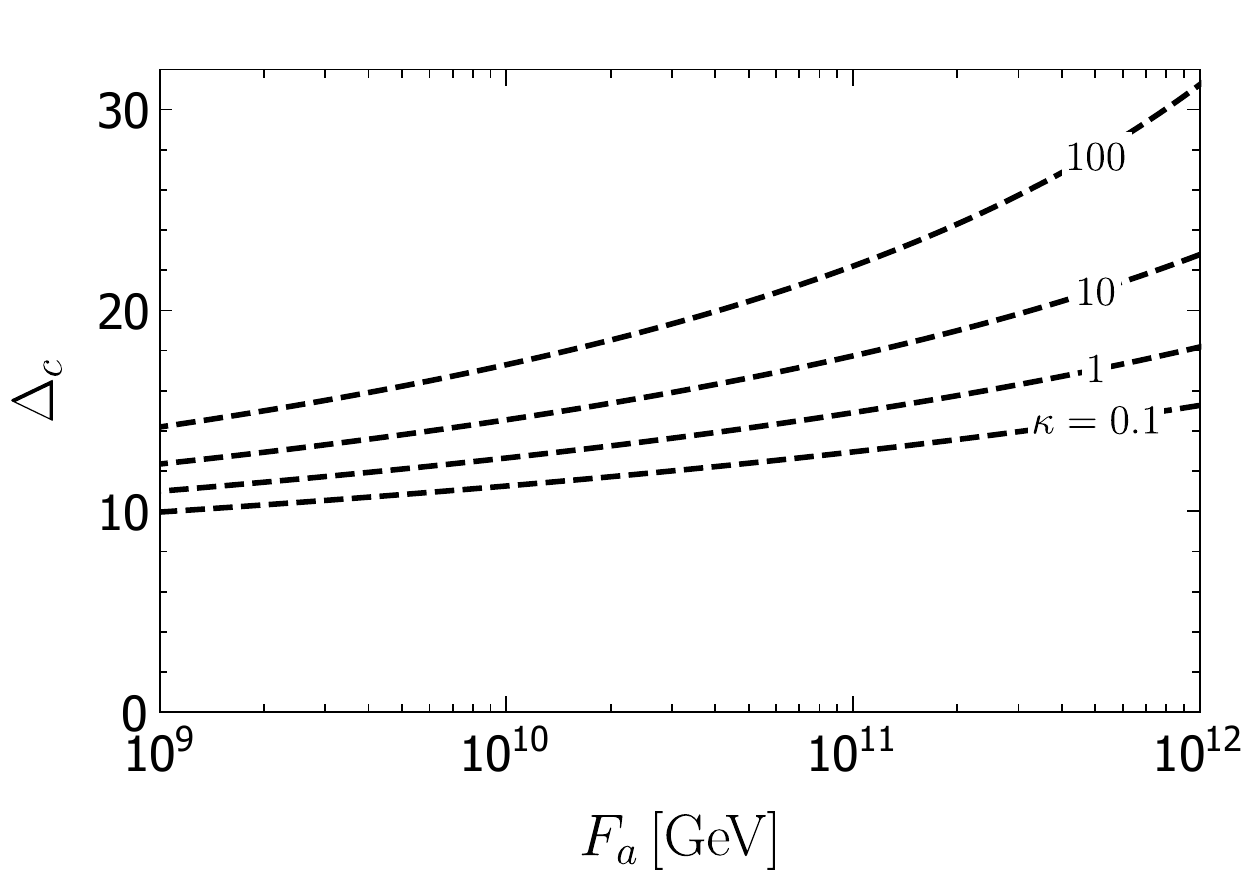}
    \caption{Left panel: QCD (solid) and UV (dashed) contributions to the axion mass for $\Delta=11$ and $\kappa=1$. Right panel: Critical operator dimension, $\Delta_c$, needed to solve the axion quality problem. We fixed $k=\,M_{P}$, $kz_{UV}=1$, $\sigma_0=0.1$, $\ell_{UV}=0.1$, and $b_{UV}=0$.}
    \label{fig:QCD}
\end{figure}

The value of $\Delta_c$ is determined by comparing the two contributions to the axion potential. In order to solve the strong CP problem the axion mass from QCD must dominate over the contribution arising from explicit UV violations of the $U(1)_{PQ}$ symmetry. The QCD instanton contribution to the axion mass is given by~\cite{diCortona:2015ldu}
\begin{equation}
    (m_a^{(QCD)})^2 \simeq \frac{m_u m_d}{(m_u+m_d)^2}\frac{m_\pi^2 F_\pi^2}{F_a^2} \simeq (5.7\,\text{meV})^2 \left(\frac{10^9\,\text{GeV}}{F_a}\right)^2 \,,
    \label{eq:QCDmass}
\end{equation}
where $m_{u,d}$ are the up, down quark masses, $m_\pi \simeq 135$\,MeV, and $F_\pi = 92$\,MeV. Combining eqs.~\eqref{eq:axion-mass-approx} and \eqref{eq:decay_constant} gives the UV contribution to the axion mass in terms of the decay constant:
\begin{equation} \label{eq:axion-mass-Fa}
  (m_a^{(UV)})^2 = \frac{4\ell_{UV}\sigma_0(\Delta-2)}{\kappa^2(\Delta-4+b_{UV})} \left( \frac{\kappa\sqrt{\Delta-1}}{\sigma_0}\right)^\Delta \left(\frac{F_a}{\Lambda_{UV}} \right)^{\Delta-4} F_a^2 \,,
\end{equation}
where we have defined $\Lambda_{UV}\equiv z_{UV}^{-1}$. The two contributions to the axion mass are shown as a function of $F_a$  in the left panel of figure~\ref{fig:QCD} for $\Delta=11$. The axion potential with both of these contributions then becomes
\begin{equation}
    V(a^{(0)}) \simeq -(m_a^{(QCD)})^2 F_a^2 \cos\left(\frac{a^{(0)}}{F_a}+\bar\theta\right) - (m_a^{(UV)})^2 F_a^2 \cos\left(\frac{a^{(0)}}{F_a}+\delta\right)\,,
    \label{eq:axionpotential}
\end{equation}
where $\delta-\bar\theta$ is the relative phase between $\bar\theta$ and the PQ-violating operator of dimension $\Delta$. The minimum of the axion potential \eqref{eq:axionpotential} is now displaced from the origin to the value
\begin{equation}
    |\bar\theta_{eff}|\equiv \bigg|\bigg\langle \frac{a^{(0)}}{F_a}+\bar\theta \bigg\rangle\bigg| \simeq \frac{(m_a^{(UV)})^2 \sin(\delta-\bar\theta)}{(m_a^{(QCD)})^2+(m_a^{(UV)})^2\cos(\delta-\bar\theta)}\,.
\end{equation}
Requiring that this shift be no larger than the experimental upper bound, $|\bar{\theta}_{eff}| \lesssim 10^{-10}$, leads to the condition $(m_a^{(UV)})^2 \lesssim 10^{-10}\, (m_a^{(QCD)})^2$, assuming an order one phase difference $\delta-\bar\theta$. This condition gives a lower bound, $\Delta_c$, on the operator dimension, with the value needed to address the axion quality problem for a given decay constant shown in the right panel of figure~\ref{fig:QCD}. The critical dimension is shown for several choices of $\kappa$, although it is mostly sensitive to just the ratio $\kappa/\sigma_0$, as can be seen from eq.~\eqref{eq:axion-mass-Fa}. For the range of decay constants $10^9~{\rm GeV}\lesssim F_a \lesssim 10^{12}~{\rm GeV}$, consistent with obtaining the relic dark matter abundance with an order one initial misalignment angle, one requires at least $\Delta_c\gtrsim 10$. 

Finally, in this class of models, the presence of $SU(3)_c$ as a 5D gauge symmetry in the bulk means that in the dual theory (some of) the fundamental constituents of the composite sector must be charged under QCD. This is analogous to 4D composite axion models considered in the literature~\cite{Lillard:2017cwx,Lillard:2018fdt,Gavela:2018paw}. Note that in our holographic description the operator dimension $\Delta$ is a free parameter, since the underlying explicit 4D model is not specified. However, requiring that the 5D theory is perturbative does imply that we are considering the composite model in the large-$N_c$ limit (where $N_c$ is the number of colours in the confining gauge group) via the relation $g_5^2k \sim (4\pi)^2/N_c$~\cite{Erlich:2005qh,DaRold:2005mxj}. Furthermore, the coupling of the axion to $G\tilde{G}$ is generated via the $U(1)_{PQ}$--$SU(3)_c^2$ anomaly and hence $\kappa \propto N_c$. 


\section{Conclusion}

The axion remains a favoured solution to the strong CP problem, as well as provides a candidate for the missing dark matter component of the Universe. However, the axion solution requires that the $U(1)_{PQ}$ global symmetry is preserved by quantum gravity to sufficiently high order terms in the Lagrangian. We have presented a 5D geometric solution to this axion quality problem that relies on a 5D gravitational dual description of composite axion models. The spontaneous breaking of a global $U(1)_{PQ}$ symmetry by a PQ-charged composite operator of dimension $\Delta$ is modelled by the vacuum expectation value of a bulk complex scalar field charged under a 5D $U(1)_{PQ}$ gauge symmetry.
The IR brane scale is associated with the scale of spontaneous $U(1)_{PQ}$ symmetry breaking, and therefore offers an explanation as to why the axion decay constant $F_a$ is much below the UV scale, while on the other hand explicit sources of global $U(1)_{PQ}$ violation are confined to the UV brane.

In our 5D setup, the axion quality problem is solved by localising the axion towards the IR brane and away from the explicit sources of $U(1)_{PQ}$ violation on the UV brane. The axion zero mode profile is controlled by the bulk scalar mass-squared parameter, which by the AdS/CFT dictionary is related to the operator dimension $\Delta$. In fact, our solution for the 5D axion profile generalises previous QCD pseudoscalar solutions~\cite{DaRold:2005vr} to operators of arbitrary dimension $\Delta$. As $\Delta$ is increased the axion becomes more IR-localised, and there is a minimum critical value $\Delta_c$, shown in figure~\ref{fig:QCD}, for which the UV contributions to the axion mass are sufficiently suppressed relative to the QCD instanton contribution, thereby preserving the solution to the strong CP problem. This requires that QCD is a gauge symmetry in the bulk, or equivalently that the composite sector is also charged under QCD. If QCD is instead confined to the UV brane, the axion cannot be sequestered from additional UV sources of explicit $U(1)_{PQ}$ violation while simultaneously maintaining a large coupling to QCD.

Our 5D geometric solution holographically captures a whole class of 4D composite axion models where the $U(1)_{PQ}$ symmetry is an accidental global symmetry of the underlying 4D strong dynamics. This is analogous to the SM where baryon number is an accidental global symmetry up to dimension six operators. If a similar mechanism were to occur for the underlying 4D gauge theory responsible for a composite axion, then our analysis suggests that the $U(1)_{PQ}$ global symmetry must be preserved up to at least dimension ten for an axion decay constant $F_a \gtrsim 10^9$\,GeV. A recent attempt to construct such a 4D model is given in \cite{Gavela:2018paw}, and other constructions with larger gauge groups should also be possible. In fact, our 5D framework can be used to model and give holographic descriptions of 4D strong dynamics with larger global symmetry groups or to consider more general possibilities with SM fermions propagating in the bulk. The 5D pseudoscalar solution may also have applications for other global symmetries, such as in QCD where chiral symmetry is broken by operators of dimension three. Thus this simple, 5D geometric solution provides a new way to understand composite axion models, and in general to study accidental global symmetries of 4D strong dynamics.


\acknowledgments
We thank Alex Pomarol for helpful discussions. The work of P.C. is supported by the Australian Research Council, and the World Premier International Research Center Initiative (WPI), MEXT, Japan. The work of T.G. and M.D.N. is supported in part by the DOE Grant No. DE-SC0011842 at the University of Minnesota, and T.G is also supported in part by the Simons Foundation. T.G. acknowledges the Munich Institute for Astro- and Particle Physics (MIAPP) of the DFG Excellence Cluster Origins, and the Kavli Institute of Theoretical Physics in Santa Barbara where part of this work was done, and also thanks the Ecole Polytechnique in Paris for hospitality and financial support while this work was being completed.


\bibliographystyle{JHEP}
\bibliography{holographic_axion.bib}

\end{document}